\documentclass[useAMS]{mn2e}
\usepackage{times}
\usepackage{epsfig}

\newcommand{\mdot}{\mbox{$\dot{M}$}}
\newcommand{\vinf}{\mbox{$v_{\infty}$}}

\def \etal   {\hbox{et~al.\/}}
\def \kms    {km~s$^{-1}$}
\def \msunyr   {{$M_{\odot}$~yr$^{-1}$}}
\def\lesssim{\mathrel{\hbox{\rlap{\hbox{\lower4pt\hbox{$\sim$}}}\hbox{$<$}}}}
\def\gtrsim{\mathrel{\hbox{\rlap{\hbox{\lower4pt\hbox{$\sim$}}}\hbox{$>$}}}}

\title[Forbidden lines from axisymmetric winds]
{Models of Forbidden Line Emission Profiles from Axisymmetric
Stellar Winds}

\author[R. Ignace and A. Brimeyer]
{
R.~Ignace and A.~Brimeyer\\
       Department of Physics, Astronomy, \& Geology,
       East Tennessee State University,
       Box 70652,
       Johnson City, TN 37614
       USA
}

\begin{document}
\maketitle


\begin{abstract}

A number of strong infrared forbidden lines have been observed in several
evolved Wolf-Rayet star winds, and these are important for deriving
metal abundances and testing stellar evolution models.  In addition,
because these optically thin lines form at large radius in the wind,
their resolved profiles carry an imprint of the asymptotic structure of
the wind flow.  This work presents model forbidden line profile shapes
formed in axisymmetric winds.  It is well-known that an optically thin
emission line formed in a spherical wind expanding at constant velocity
yields a flat-topped emission profile shape.  Simulated forbidden
lines are produced for a model stellar wind with an axisymmetric density
distribution that treats the latitudinal ionization self-consistently and
examines the influence of the ion stage on the profile shape. The
resulting line profiles are symmetric about line centre.  Within a
given atomic species, profile shapes can vary between centrally peaked,
doubly peaked, and approximately flat-topped in appearance depending on
the ion stage (relative to the dominant ion) and viewing inclination.
Although application to Wolf-Rayet star winds is emphasized, the concepts
are also relevant to other classes of hot stars such as luminous blue
variables and Be/B[e] stars.

\end{abstract}

\begin{keywords}
stars: emission line, Be -- stars: mass-loss --
stars:  winds, outflows -- stars: Wolf-Rayet 
\end{keywords}

\section{Introduction}

Outstanding questions about the three-dimensional structure of stellar
winds continue to plague our understanding of mass-loss processes from
stars.  Although much has been accomplished toward understanding the gross
properties of stellar winds in terms of theories for spherically symmetric
mass-loss (a description of current understanding along with copious
references in this regard can be found in Lamers \& Cassinelli 1999),
a complete physical description of two and three dimensional effects remains
elusive.  There have been notable advances in a number of areas, with
a brief and incomplete listing to include such studies as axisymmetric
models of magnetized thermally driven winds (Sakurai 1985; Washimi \&
Shibata 1993; Sean \& Balick 2004), axisymmetric line-driven winds with
rotation (e.g., Bjorkman \& Cassinelli 1993; Owocki, Cranmer, \& Gayley
1996; Petrenz \& Puls 2000), axisymmetric magnetized line-driven winds
(Poe, Friend, \& Cassinelli 1989; Brown \etal\ 2004; Ud-Doula \& Owocki
2002; Townsend \& Owocki 2005), axisymmetric dust-driven winds (Asida \&
Tuchman 1995; Dorfi \& H\"{o}fner 1996; Soker 2000), and time-dependent
spherical or time-averaged spherical models (Feldmeier 1995; Dessart \&
Owocki 2002; Runacres \& Owocki 2002, 2005). The challenge has been that
to include all of the relevant physics could require multi-dimensional
radiative (magneto-)hydrodynamics.  Despite the difficulties involved,
semi-analytic and detailed numerical models have been inspired by a
large volume of observations that exhibit behavior that the steady-state
spherical models are unable to explain.  The examples are numerous,
such as the triple rings of the SN1987A remnant (Crotts, Kunkel, \&
McCarthy 1989; Burrows \etal\ 1995; Sugerman \etal\ 2005), the fact
that the majority of planetary nebulae and luminous blue variables are
aspherical (Balick 1987; Nota \etal\ 1995; Davies, Oudmaijer, \& Vink
2005), the existence of discs in the Be and B[e] stars (see the review of
Be stars by Porter \& Rivinius 2003; see the recent proceedings on B[e]
stars by Kraus \& Miroshnichenko 2006), discrete absorption components
and related phenomena seen in hot star wind lines (Howarth \& Prinja 1989;
Massa \etal\ 1995; Brown \etal\ 1995; Lepine, Moffat, \& Henriksen 1996),
and the inference of apparently ubiquitous clumpiness in stellar winds
(Hillier 1991; Bouret, Lanz, \& Hillier 2005; Fullerton, Massa, \&
Prinja 2006).

The goal of this contribution is to consider diagnostics of aspherical
mass-loss as revealed by forbidden emission line profiles.  The
formation of most forbidden lines occurs over a large-scale volume
of the wind.  Consequently, the emission profile shapes that form
in the hypersonic winds of early-type stars are imprinted with the
properties of the asymptotic wind structure.  This is most certainly
not a new concept.  Dating back to Beals (1929), it was appreciated
that flat-topped emission profile shapes result for optically thin
lines that form in a constant velocity and radially expanding flow.
In a colliding wind binary, Stevens \& Howarth (1999), L\"{u}hrs
(1997), Hill \etal\ (2000), and Hill, Moffat, \& St-Louis (2002)
have shown how time-dependent deviations from flat-top shaped
profiles can be used as a tracer of the geometry of the colliding
winds.

In this paper a parametric axisymmetric density distribution is adopted
as a framework for modeling forbidden emission profiles as a function of
the wind asymmetry and viewing inclination.  In addition, we point out
that the density function alone is not enough to completely determine
emission profile shapes.  The emissivity depends on the ionic species
and its latitude dependence that is is coupled to the overall density
distribution.  So our model profiles also include a treatment of the
influence of latitudinal ionization effects.  Interestingly, for a given
stellar wind model, different lines can show different, even opposite,
profile trends, as for example some lines show centrally-peaked emission,
whereas others display double-peaked profiles.  A number of simplifying
assumptions are employed to focus on the general trends.  The wind
expansion is assumed constant for large radius and also in latitude.
Clumping is ignored.  How clumping may vary with latitude and radius in
an axisymmetric hot star wind has not been well-studied, and so we choose
to avoid making {\it ad hoc} assumptions.  However, the basic equations
presented here can be modified to include velocity and clumping effects
to study their impact on forbidden line profile shapes.

Section 2 presents a brief overview of forbidden line formation in
spherically symmetric stellar winds.  Then Section 3 contains a
description of the axisymmetric model, and results for model forbidden
line profiles are given.  Following is a summary and discussion of
applications in Section~4.  There are two appendices to conclude
the paper.  Appendix~\ref{poleon} presents an analytic solution for
line profile shapes in the case of a pole-on viewing perspective
of an axisymmetric wind, and Appendix~\ref{2lvl} specifically shows
that the two-level atom approximation that will be adopted throughout
the paper applies to the particular and interesting case of Ne{\sc
iii} for Wolf-Rayet star conditions.

\section{Formation of Forbidden Lines in Spherical Winds}

Our presentation of the theory of forbidden line emission generally
follows the approach and to some extent the notation of Osterbrock
(1989) regarding the atomic physics and Barlow, Roche, \& Aitken (1988)
in terms of stellar wind application.

\subsection{The Atomic Physics}

Consider a fine structure transition for an ion
species assuming a two-level atom only, denoting `1' for the lower
level and `2' for the upper level.  The level population density
(number per unit volume) $n_1$ can be modified by collisional and
radiative excitations, and the level population density $n_2$ can be
modified by collisional de-excitations and spontaneous radiative decays.
Highly non-LTE conditions exist in hot stellar winds, and so the two-level
atom representation can be an excellent approximation in many cases.
However, it can be the case for some atoms that there are more than
two levels that can be important for the population that determines
the emission for the line of interest.  In the case of the
Wolf-Rayet (WR) stars, the lines of [Ne{\sc ii}] 12.8 $\mu$m ($^2P_{1/2}-
^2P_{3/2}$) and [Ne{\sc iii}] 15.56 $\mu$m ($^3P_{1}- ^3P_{2}$) are of
considerable interest for determining the abundance of neon.
Although, the ground state of Ne{\sc ii} is a doublet and so should
be well-described by the two-level atom approximation, Ne{\sc iii} has
five low-lying levels.  The particular line at 15.56 $\mu$m happens to
involve the two lowest levels of the five, and we will show later that
the two-level approximation is pretty good for conditions relevant to
WR stars.  Thus the theory that follows assumes a two-level atom throughout. 

It is commonly the case that radiative excitations from 1 to 2 are
ignored compared to the collisional term.  The radiative excitation
rate (number per volume per second) is $n_1 \, B_{12}\, J_\nu$, with
$B_{12}$ the upward Einstein transition probability, and $J_\nu$ the mean
intensity at the frequency of the line transition.  The collisional rate
is $n_1\,n_{\rm e}\, q_{12}$, where $n_{\rm e}$ is the electron number
density and $q_{12}$ is the collisional excitation rate 
in volume per second.  A ratio of these rates shows that the
radiative excitation rate is indeed much smaller than the collisional
rate.  For the wind case, $J_\nu$ will decrease with radius in the wind
as the dilution factor that scales as $r^{-2}$, and this cancels with the
electron density in the denominator that will also scale in the same way.
The dominant factor that remains is a ratio $n_{\rm c}/n_{\rm e,0}$,
where $n_{\rm c}$ is the `critical density' and $n_{\rm e,0}$ (to be
defined in the following section) is a scale factor for the electron
number density.  The critical density is defined here as

\begin{equation}
n_{\rm c} = \frac{A_{21}}{q_{21}}.
	\label{eq:crit}
\end{equation}

\noindent In effect, the critical density represents a transition
between high density regimes with $(n_{\rm e} \gg n_{\rm c})$ for
the line emissivity is linear in density versus low density regimes
with $(n_{\rm e} \ll n_{\rm c})$ for which the emissivity scales
with the square of density.

The volume emissivity $j$ (erg s$^{-1}$ cm$^{-3}$ sr$^{-1}$) is given by

\begin{equation}
j = \frac{1}{4\pi}\,h \, \nu_{21} \, A_{21}\, n_2.
	\label{eq:emiss}
\end{equation}

\noindent To determine $n_2$, equilibrium conditions for the two levels are 
imposed as governed by the
following rate expression:

\begin{equation}
n_1 \, n_{\rm e} \, q_{12} = n_2\, n_{\rm e} \, q_{21} + 
	n_2\, A_{21}, 
	\label{eq:bal}
\end{equation}

\noindent where

\begin{equation}
\frac{q_{21}}{q_{12}} = \frac{g_1}{g_2}\,e^\beta,
\end{equation}

\noindent with 

\begin{equation}
\beta = h\nu_{21}/kT_{\rm e}, 
\end{equation}

\noindent for $\nu_{21}$ the frequency of the line transition, $T_{\rm e}$
the electron temperature, and $g$ the statistical weight of the level,
with $g=2J+1$.  The collisional volume rate is given by 

\begin{equation}
q_{21} = \frac{8.629\times10^{-6}}{T_{\rm e}^{1/2}}\,\frac{\Omega_{12}}{g_2},
\end{equation}

\noindent where $\Omega_{12}$ is the collision strength.

We define $n_{\rm i,E}$ to be the number density of element $E$ in ion
stage $i$, and we approximate its value as

\begin{equation}
n_{\rm i,E} \approx n_1 + n_2.
	\label{eq:ground}
\end{equation}

\noindent We make the following definitions: the ion fraction for this
element in this ion stage is

\begin{equation}
Q_{\rm i,E} = \frac{n_{\rm i,E}}{\sum_{\rm i}\,n_{\rm i,E}} \equiv
	\frac{n_{\rm i,E}}{n_{\rm E}};
\end{equation}

\noindent the abundance of the element relative to all nucleons is

\begin{equation}
{\cal A}_{\rm E} = \frac{n_{\rm E}}{n_{\rm N}},
\end{equation}

\noindent for $n_{\rm N}$ the number density of nucleons; and
the ratio of the nucleons to electrons is denoted

\begin{equation}
\gamma_{\rm e} = \frac{n_{\rm N}}{n_{\rm e}}.
\end{equation}

\noindent With these relations, the emissivity in
equation~(\ref{eq:emiss}) can be re-expressed in terms of $n_{\rm e}$ 
instead of $n_2$.  First, equation~(\ref{eq:ground}) is used to eliminate
$n_1$ in equation~(\ref{eq:bal}), the result of which can be solved for
$n_2$ alone, giving

\begin{equation}
n_2 = \frac{n_{\rm i,E}\,n_{\rm e}\,q_{12}}{n_{\rm e}\,q_{12}
	+n_{\rm e}\,q_{21}+A_{21}}.
\end{equation}

\noindent Factoring out $n_{\rm e} q_{12}$, using the definition of the
critical density from equation~(\ref{eq:crit}) in conjunction with the
preceding relations, we arrive at

\begin{equation}
n_2 = Q_{\rm i,E}\,{\cal A}_{\rm E}\,\gamma_{\rm e}\,n_{\rm e}\,
	\left\{1 + \frac{g_1}{g_2}\,e^\beta+\frac{g_1\,n_{\rm c}}
	{g_2\,n_{\rm e}}\,e^\beta\right\}^{-1}
	\label{eq:n2}
\end{equation}

\noindent At high densities for which $n_{\rm e}\gg n_{\rm c}$, one
has that $j \propto A_{21}\,n_{\rm e}$, whereas at low densities with
$n_{\rm e}\ll n_{\rm c}$, the scaling is $j \propto A_{21}\,n_{\rm
e}^2/n_{\rm c} \propto q_{12}\,n_{\rm e}^2$.  

In Barlow \etal\ (1988), the quantity $\gamma_{\rm i}$ is
defined as the {\it ionic fraction}, being the number density ratio of
a particular ion of a particular atom to all ions in the gas.  In our
notation, one has that $\gamma_{\rm i} = Q_{\rm i,E}\,{\cal A}_{\rm E}$.
It is desirable to derive the ionic fraction to compare against stellar
evolution predictions.  The ionic fraction can be inferred from the observed
total flux of forbidden line emission, provided that the source distance
is known.  If lines from different atomic species are available, line
ratios can be used to derive abundances (or possibly limits to abundances)
even if the source distance is not well-known.

\begin{figure}
\centerline{\epsfig{file=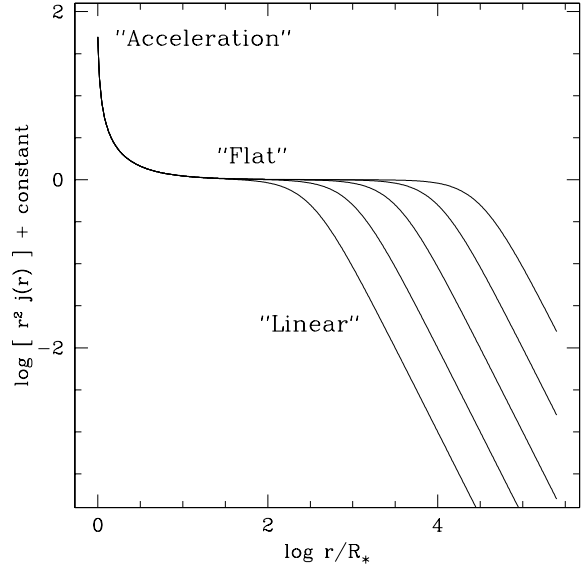,angle=0,width=8cm}}
\caption[]{A logarithmic plot of the radial gradient of the line
luminosity, $dL_l/dr \propto r^2 j(r)$, versus normalized radius in
the wind.  The curve shows three main components.  The first is the
`Acceleration' component at far left where $v(r)$ changes from some low
value at the wind base to $\vinf$.  Middle is the `Flat' portion, where
$r^2j \propto constant$.  Right is the `Linear' part where $r^2j \propto
r^{-2}$. The vertical scale is shifted so that the `Flat' portion is at
zero.  The different breaks from the flat segment to the linear portions
represent line transitions with different critical densities (see text).
\label{fig1}}
\end{figure}

\subsection{Spherical Winds}

It is instructive to briefly review the formation of forbidden lines in a
spherical wind.  For this purpose it is useful first to develop a sense
of scale, to determine where the line emission is formed in the wind flow.

Consider a spherical wind with mass-loss rate \mdot\ and radial speed
$v(r)$ for radius $r$.  The mass density will be

\begin{equation}
\rho_{\rm sph} = \frac{\mdot}{4\pi\,r^2\,v(r)}.
\end{equation}

\noindent The number density will be $n_{\rm sph} = n_{\rm N} + n_{\rm e}
= \rho_{\rm sph} / \bar{m}$, with $\bar{m}=\mu m_H$ the average particle
mass in the wind, and $\mu$ the mean molecular weight.  The electron
number density is $n_{\rm e} = \rho_{\rm sph}/\mu_{\rm e}m_H$, for
$\mu_{\rm e}$ the mean molecular weight per free electron.  Similarly,
the nucleon number density is $n_{\rm N} = \rho_{\rm sph}/\mu_{\rm N}m_H$,
for $\mu_{\rm N}$ the mean molecular weight per free nucleon.  In a hot
star wind that is dominated by ionized hydrogen, the relations become
$\bar{m} = m_{\rm H}/2$, $\mu_{\rm e} = \mu_{\rm N} = 1$, $n_{\rm e}
= n_{\rm N}$ and so $\gamma_{\rm e} =1$.  This would be appropriate
for an O star, for example.  Strong forbidden lines are seen in the IR
spectra of WR stars, and a WR~wind is often dominated by twice ionized
helium at small radii but possibly once ionized helium at large radii
(e.g., Fig.~3 of Dessart \etal\ 2000 shows He{\sc ii} is dominant for
their WC8 model of WR~135).  Assuming that helium is once ionized,
the molecular weights become $\mu = 2$, $\mu_{\rm e} = 4$,
and $\mu_{\rm N} = 4$, and $\gamma_{\rm e} = 1$.

The region of line formation is roughly set by the radius at which the
electron density and the critical density are equal.  Among hot stars the
mass loss rates are around $10^{-10}$ \msunyr\ and larger, with terminal
speeds generally around 1000 \kms\ or greater from spectral type B and
earlier (e.g., Lamers \& Cassinelli 1999).  We define a number density
scale factor to be

\begin{equation}
n_0 = \frac{\mdot}{4\pi\,R_*^2\,\vinf\,\mu\,m_H},
\end{equation}

\noindent with values from $10^7$ cm$^{-3}$ for B star winds
soaring to $10^{14}$ cm$^{-3}$
in the case of dense WR winds.  The critical density for a forbidden
line depends on the electron number density, so we define a related
scaling factor for the electron density:

\begin{equation}
n_{\rm e,0} = \mu \, n_0 / \mu_{\rm e}.
\end{equation}

\noindent Typical stellar wind forbidden lines observed in WR winds 
have critical densities ranging from $10^4$ to $10^7$ cm$^{-3}$
(e.g., see tab.~5 of Ignace \etal\ 2001).  Ignoring the weaker winds
with quite small \mdot\ values, the radius in the wind where $n_{\rm e} =
n_{\rm c}$ occurs is expected to be very large.  We can find this radius
$r_{\rm c}$ from $n_{\rm e} = (n_{\rm e,0})\, (R_*^2/r_{\rm c}^2) =
n_{\rm c}$, giving

\begin{equation}
\frac{r_{\rm c}}{R_*} = \sqrt{\frac{n_{\rm e,0}}{n_{\rm c}}},
\end{equation}

\noindent with typical values of $10^2 - 10^4$ stellar radii.

A goal of studying forbidden lines in stellar winds is to derive
information about metal abundances.  The observable is the total line
flux that is related to the line luminosity as $F_l = L_l /4\pi D^2$,
where $D$ is the distance to the source.  Being an optically thin line,
the total line luminosity depends on the atomic rates associated with the
line transition and on the wind parameters, including the ionic fraction.
With the atomic rates known, and the wind properties such as mass-loss
rate and terminal speed provided, the observed line emission can be used
to find the ionic fraction.

The total luminosity of the optically thin line emission
is given by a volume integration involving the emissivity, viz

\begin{equation}
L_l = 4\pi\,D^2\,F_l = 4\pi\,\int_{R_*}^\infty\,4\pi\,r^2\,j(r)\,dr.
	\label{eq:linelum}
\end{equation}

\noindent Defining $\omega = (g_1/g_2)\,e^\beta$, then substituting 
equation~(\ref{eq:n2}) into equation~(\ref{eq:emiss}), and this
again into equation~(\ref{eq:linelum}), one obtains 

\begin{equation}
L_l = L_0\,\int_{R_*}^\infty\,\frac{dr/R_*}{1+\omega+\omega\,(r/r_{\rm c})^2},
\end{equation}

\noindent where

\begin{equation}
L_0 = 4\pi\,Q_{\rm i,E}\,{\cal A}_{\rm E}\,\gamma_{\rm e}\,h\,\nu_{21}\,
	A_{21}\,n_{\rm e,0}\,R_*^3.
	\label{eq:L0}
\end{equation}

Using a change of variable with $u=R_*/r$, the integral can be
re-expressed as

\begin{equation}
L_l = L_0\, \int_0^1\,\frac{du}{(1+\omega)\,u^2 + \omega\,u_{\rm c}^2},
\end{equation}

\noindent where $u_{\rm c} = R_*/r_{\rm c}$.  This has an analytic
solution:

\begin{equation}
L_l = L_0\,\frac{1}{\sqrt{(1+\omega)\,\omega\,u_{\rm c}^2}}\,\tan^{-1}\,
	\sqrt{\frac{1+\omega}{\omega\,u_{\rm c}^2}}.
\end{equation}

\noindent With $\omega$ a factor of order unity, and the expectation that
$u_{\rm c}^2 \ll 1$, the total luminosity generated for this forbidden
line is 

\begin{equation}
L_l \approx L_0\,\frac{\pi/2}{\sqrt{(1+\omega)\,\omega}}\,\frac{r_{\rm c}}{R_*}
	\propto A_{21}^{1/2}\,n_{\rm e,0}^{3/2},
	\label{eq:Ll}
\end{equation}

\noindent which is in effect the same as equation~(13) of Smith \& Houck
(2005), except that those authors allow for a constant clumping factor
(see our discussion of clumping in sect.~\ref{models}).  From 
equation~(\ref{eq:Ll}) one can
easily backsubstitute to obtain the product of the ion fraction and metal
abundance in terms of the measured line luminosity (for known distance):

\begin{equation}
Q_{\rm i,E}\,{\cal A}_{\rm E} = \frac{(1+\omega)^{1/2}\,\omega^{1/2}\,n_{\rm c}^{1/2}\,L_l}
	{2\pi^2\,\gamma_{\rm e}\,h\,\nu_{21}\,A_{21}\,n_{\rm e,0}^{3/2}\,R_*^3}
\end{equation}

The reader may object that our derivation is lacking an important
component of the wind, namely the inner radii where the wind accelerates.
We have so far tacitly assumed a constant expansion wind at $v=\vinf$
to derive the above result.  At the inner wind, the flow accelerates
from some near-hydrostatic configuration at subsonic speeds to the
hypersonic speeds of order $10^3$ \kms.  For a typical forbidden line
and a fairly dense O or WR star wind, the value of the critical radius
is so large, that the contribution to the line flux from the very dense
inner and accelerating wind region is negligible.  To illustrate this,
Fig.~\ref{fig1} shows the contribution function for the line luminosity.
Plotted is $dL_l/dr \propto j r^2$ representing the line emission from a
spherical shell as a function of the radius.  A wind velocity law with
the form $v(r) = \vinf\, (1-b/r)$ is assumed (with $b$ defined so that
$v=v_0$ some initial wind speed at radius $r=R_*$), such that $n_0 =
10^{13}$ cm$^{-3}$, and $n(R_*) \approx 10^{15}$ cm$^{-3}$.

At small radii, the curve is sharply dropping because of the rapidly
changing velocity law.  This portion is labelled the `Acceleration'
branch.  Interior to the critical radius (where $n_{\rm e} \gg n_{\rm
c}$), the contribution function is constant, and this branch is labelled
as `Flat'.  At large radius beyond the critical radius, the emissivity
drops as the square of density, where $r^2j \propto n_{\rm e}$, and so
is labelled `Linear'.  The different
breaks from the flat portion are representative of forbidden lines
with different critical densities, here ranging from $10^4$ cm$^{-3}$
(break at largest radius) to $10^8$ cm$^{-3}$ (break at smallest radius).

Although the Acceleration branch shows quite a strong contribution
function, it is relegated to such a narrow range of radii as compared
to the location of the critical radius, that it can safely be ignored
for computing the emission line flux and profile shape.  What remains
is a constant velocity spherical expansion flow that forms a flat-topped
emission profile, which is the standard theoretical result for this case
of an optically thin line (e.g., Mihalas 1978; Lamers \& Cassinelli 1999).
Note that the Flat and Linear portions contribute approximately equally
to the total line luminosity.

\section{Axisymmetric Winds}

\subsection{Density Structure}

With a good picture of the forbidden line formation in spherical winds
in mind, the case of axisymmetric flows is now addressed.  To model such
winds, the following parametric forms for the wind density are adopted.
We assume the wind is top-bottom symmetric in what follows.
Then if the equator has higher density than the pole, we use

\begin{equation}
n(r,\theta_*) = n_{\rm p}(r)\,\left[ 1 + (G-1)\,\sin^m \theta_*\right],
	\label{eq:neq}
\end{equation}

\noindent where the star is taken to have spherical coordinates $(r,
\theta_*, \phi_*)$, $n_{\rm p}(r)$ is the purely radius-dependent number
density distribution along the polar axis, and $G = n_{\rm eq}/n_{\rm
p}\ge 1$ is the equator-to-pole density contrast.  The latitude dependence
of the density is modeled as a sine function of the co-latitude $\theta_*$
to some power $m$.  This density function is motivated by examples of
equatorial density enhancements inferred for some classes or particular
stars, such as B[e] stars, or WR~134 that will be discussed later.
In the case of polar enhanced wind flows, $G = n_{\rm p} / n_{\rm eq}
\ge 1$, and we use

\begin{equation}
n(r,\theta_*) = n_{\rm eq}(r)\,\left[ 1 + (G-1)\,\cos^m \theta_*\right],
	\label{eq:np}
\end{equation}

\noindent models that are inspired by systems like $\eta$ Car and other
Luminous Blue Variables that show bipolar morphologies.  These density
functions are not specifically related to any particular model, but are
chosen as a convenient way of characterizing axisymmetric winds.

To characterize the degree of asymmetry in the wind, an opening angle
$\Delta \theta$ refers to the latitude interval between either the pole
or equator and the latitude at which $n(r,\theta_*) = 0.5\, (n_{\rm p}
+n_{\rm eq})$.  This works out in value to be the same for both polar
or equatorially enhanced winds, as given by

\begin{equation}
\Delta \theta = \cos^{-1}\,\left(2^{-1/m}\right).
	\label{eq:angle}
\end{equation}

\begin{figure}
\centerline{\epsfig{file=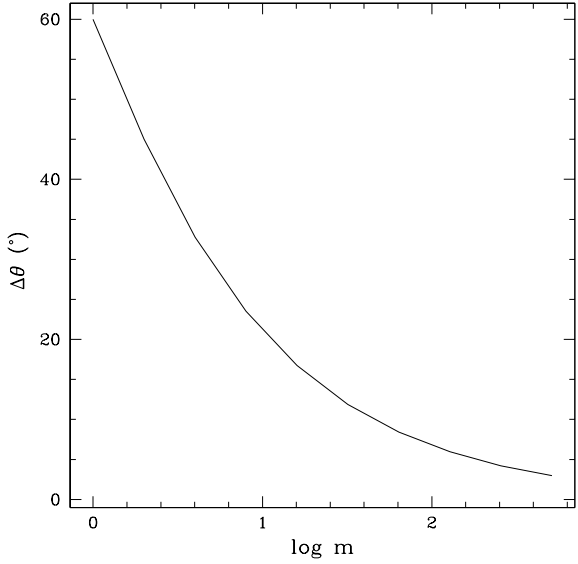,angle=0,width=8cm}}
\caption[]{
A plot of the effective opening angle $\Delta \theta$ of a polar or 
equatorial wind plotted as a function of the power exponent $m$.
\label{fig2}}
\end{figure}

\noindent Values of $\Delta \theta$ are plotted
in Fig.~\ref{fig2} against the logarithm of the power exponent
$m$.  Within this parametrization, well-confined polar or equatorial
flows require quite large values of $m$.  For example, in order to
achieve $\Delta \theta \approx 5^\circ$, a value of $m \sim 10^2$
is required.

With forbidden line emission formed over a large-scale volume, we
shall ignore the inner wind acceleration based on the arguments of
the previous section, and treat all densities to vary as $r^{-2}$.
To compare the spherical outflow case with the axisymmetric
models, the following normalization is used:

\begin{equation}
n_{\rm sph}(r) = \frac{1}{2}\,\int_{-1}^{+1}\,n(r,\theta_*)\,d\mu_*,
	\label{eq:norm}
\end{equation}

\noindent where $\mu_* = \cos \theta_*$.  For a given value of $G$
and $m$, this equation determines the ratio $H_{\rm eq} = n_{\rm
sph}(r)/n_{\rm p}(r)$ for a dense equatorial wind or $H_{\rm p} = n_{\rm
sph}(r)/n_{\rm eq}(r)$ for a dense polar wind.

\begin{figure}
\centerline{\epsfig{file=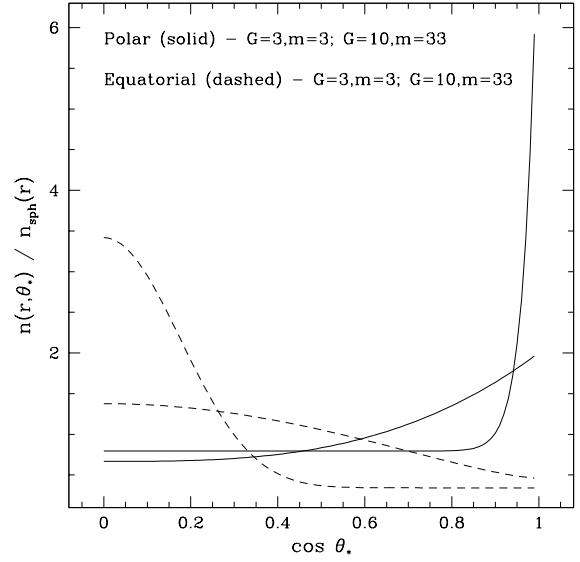,angle=0,width=8cm}}
\caption[]{
A plot of the latitude-dependent density normalized to the density that
a spherical wind would have against stellar co-latitude.  The curves are
for polar enhanced (solid) and equatorially enhanced (dashed) winds for
values of $G=3$ and $m=3$, or $G=10$ and $m=33$.
\label{fig3}}
\end{figure}

To illustrate, example density profiles are plotted in Fig.~\ref{fig3}
for different values of $G$ and $m$.  The density is normalized to
the spherical equivalent, namely a spherical wind with the same
mass-flux as the axisymmetric one.  The density is plotted as a
function of stellar co-latitude in $\cos \theta_*$, and so the
equator is at left, and the pole at right.  The solid curves are
for the polar enhanced winds, and the dashed for denser equatorial
winds.  A brief comment on the dense polar winds versus the dense
equatorial winds.  When $m$ and $G$ are constant, $n_{\rm p}$ will
be larger for a dense polar wind than the corresponding value of
$n_{\rm eq}$ for a dense equatorial wind.  Although the opening
angle $\Delta \theta$ will be the same, the solid angle associated
with the respective dense components is not equal, being smaller
in the case of a polar flow.

\subsection{Ionization Structure}

For the ionization balance in the wind, steady-state conditions are
assumed.  Only radiative recombination and photo-ionization are
considered.  The ion stage $i$ ranges from 0 to $I$, where `0'
denotes a neutral atom, and `$I$' complete ionization.  Since
interest is focussed on forbidden lines of trace metals, the electron
number density is assumed {\it not} to be impacted by the ionization
balance of the metals.  For example in O stars, hydrogen can be
safely considered as completely ionized with $n_{\rm e} = n_{\rm
H}$ for all radii and latitudes.  For a WR~wind, singly or doubly
ionized helium would dominate the electron number density (although
ionization of C and O is relevant for the carbon-rich WR stars).
Mainly modest distortions of the wind from spherical are considered,
and not dense equatorial discs or polar jets.  The simplification
is that with $n_{\rm e}$ given, the ionization balance for different
metal species is independent.

Under these conditions, ionization equilibrium will be governed
by equations of the form

\begin{eqnarray}
0 & = & +\alpha_{\rm i+1}(T_{\rm e})\,n_{\rm e}\,n_{\rm i+1} - \alpha_{\rm i}
	(T_{\rm e})\,n_{\rm e}\,
	n_{\rm i} \nonumber\\
 & & +\int_{\nu_{\rm i-1}}^\infty 
	\frac{J_\nu}{h\nu}\,\sigma_\nu(i-1)\,n_{\rm i-1}\,d\nu  \nonumber\\
 & &	- \int_{\nu_{\rm i}}^\infty\,
	\frac{J_\nu}{h\nu}\,
        \sigma_\nu(i)\,n_{\rm i}\,d\nu , 
	\label{eq:ioniz}
\end{eqnarray}

\noindent where the ion stages are all for the same element so that the
subscript `E' has been dropped, $\alpha_{\rm i}$ is a recombination
coefficient for ion $i$, $J_\nu$ is a mean intensity, $\nu_{\rm i}$ is
the photo-ionization edge frequency for ion $i$, and $\sigma_\nu(i)$
is the photo-ionization cross-section for ion $i$.  In general, the
equation directly couples three adjacent ion stages; implicitly all of
the stages are coupled through the fact that $n_{\rm E} = \sum_{\rm
i}\, n_{\rm i}$.  To complete the set of equations, one has that for
the equation involving $i=0$, there is no recombination to a lower ion
stage, and for the case $i=I$, there is no photo-ionization to a higher
ion stage.

These expressions can be simplified considerably, by assuming the diffuse
radiation is negligible.  The application of the theory in this paper
emphasizes WR~winds, and it is well-known these winds are so dense that
the continuum forms in the wind itself, and a classic static photosphere
is not observed.  As such, the stellar radius is ill-defined, being
wavelength dependent.  Still we adopt a core-halo approach,
in which the continuous opacity is thin beyond the radius where the
continuum forms, because this is a good approximation for ascertaining the
ionization balance at large radii where the forbidden line emission
is generated.

To determine the ionization balance from a set of expressions
based on equation~(\ref{eq:ioniz}), it is useful to define the
following dimensionless parameter:

\begin{equation}
\delta_{\rm i} = \frac{\Gamma_{\rm i}}{n_{\rm e,0}(\theta_*)\,\alpha_{\rm i+1}},
	\label{eq:delta}
\end{equation}

\noindent where

\begin{equation}
n_{\rm e,0}(\theta_*) = \frac{\mu}{\mu_{\rm e}}\,n_0\,\left\{ \,
	\begin{array}{c} 
	H_{\rm p}^{-1} \, [ 1+(G-1)\cos^m \theta_* ]  \\ 
	{H_{\rm eq}^{-1} \, [ 1+(G-1)\sin^m \theta_* ]} 
	\end{array} 
	\right.
\end{equation}

\noindent for the appropriate dense polar or dense equatorial wind
case.  For convenience, we introduce $\delta_{-1} = 1$.  The factor
$\Gamma_{\rm i}$ pertains to the photo-ionization rate, as given
by

\begin{equation}
\Gamma_{\rm i} = \frac{1}{4}\,\int_{\nu_{\rm i}}^\infty\,
        \frac{I_\nu^*}{h\nu}\,\sigma_\nu(i)\,d\nu,
	\label{eq:Gamma}
\end{equation}

\noindent where the factor of $1/4$ comes from the dilution factor
in the large radius limit.  It is also seen that the latitude dependence for a
wind with an axisymmetric density distribution enters into the ionization
balance via the recombination rate terms.  The ion fraction
$Q_{\rm i,E}$ will be given by

\begin{equation}
Q_{\rm i,E}(\theta_*) = \frac{\Pi_{j=-1}^{i-1}\,\delta_{\rm j}}
	{\sum_{j=-1}^{I-1}\,\Pi_{k=-1}^{k=j}\,\delta_{\rm k}}.
	\label{eq:Q}
\end{equation}

\noindent This form also holds for the Saha equation,
such as found in Mihalas (1978, eq.~5.17).  However, we stress again
that equation~(\ref{eq:Q}) for the solution of the ionization balance
specifically refers to trace metals in which the electron number
density is not substantially altered by the ionization balance in
the element E under consideration.

It is useful to consider some special cases in relation to the
structure of expression~(\ref{eq:Q}).  Consider the condition in
which there is one dominant ion stage $i_0$, with $Q_{\rm i_0}
\approx 1$.  The normalizations are such that as the stellar intensity
is made either larger in amount or harder in energy, for a given base
density $n_0$, the values of $\delta_{\rm i}$ will increase.  As the
dominant ion moves to higher stages, the lower ion stage fractions will
become less significant.  

Consider the expression for the dominant stage, as given by

\begin{equation}
Q_{\rm i_0} = \frac{\delta_{-1} \,\cdots\, \delta_{\rm i_0-1}}
	{\delta_{-1} + \delta_{-1}\delta_0 + \,\cdots\, +
	(\delta_{-1}\cdots \delta_{\rm I-1})}
\end{equation}

\noindent Stages with $i>i_0$ are not dominant, which can only mean that
the photo-ionization rate is not large compared to the recombination rate,
and so $\delta_{\rm i} \ll 1$ for $i>i_0$.  Since the other
factors are large compared to unity (by definition, otherwise the stage
$i_0$ could not be dominant if lower stages were not trace), then in the
denominator, the normalization must be dominated by the term that involves
the product $\delta_{-1} ... \delta_{\rm i_0-1}$, and so clearly $Q_{\rm
i_0} \approx 1$ since the numerator and denominator have like factors.

How then will adjacent stages scale with the density?  For the next
lower stage and using the above argument for the dominant term in the
denominator, the ion fraction will be

\begin{equation}
Q_{\rm i_0-1} \approx \frac{\delta_{-1} \,\cdots\, \delta_{\rm i_0-2}}
	{\delta_{-1} \,\cdots\, \delta_{\rm i_0-1}} = \frac{1}
	{\delta_{\rm i_0-1}}\propto n_0(\theta_*).
\end{equation}

\noindent Similarly, the next higher stage will scale as

\begin{equation}
Q_{\rm i_0+1} \approx \frac{\delta_{-1} \,\cdots\, \delta_{\rm i_0}}
        {\delta_{-1} \,\cdots\, \delta_{\rm i_0-1}} = 
        \delta_{\rm i_0}\propto \left[n_0(\theta_*)\right]^{-1}.
\end{equation}

\noindent Thus we arrive at the general rule, in the case where there is
clearly a single dominant ion stage, that

\begin{equation}
Q_{\rm i} \propto \left[n_0(\theta_*)\right]^{i_0-i}.
	\label{eq:scale}
\end{equation}

\noindent This is not an especially fresh revelation.  For example,
the result of equation~(\ref{eq:scale}) with its dependence on
$i_0-i$ for the case of spherical winds can be found in spirit in the
discussion of Lamers, Cerruti-Sola, \& Perinotto (1987).  (The scaling
appears more clearly in Bjorkman \etal\ 1994 in their eq.~[17].)

A latitude-dependent ionization balance is important for the
bi-stability effect (Pauldrach \& Puls 1990) for generating
axisymmetric winds from the influence of stellar rotation (Lamers
\& Pauldrach 1991), although in this case it is a combination of a
latitude-dependent stellar radiation field, mass-loss rate, and
terminal speed that generates the axisymmetric wind density, whereas
we are concerned with a `generic' latitude-dependent density and
its feedback on the ionization balance in the wind.

The radiative hydrodynamic models of Petrenz \& Puls (2000), who
modeled axisymmetric line-driven winds from rotating stars, provide
ion fractions with latitude (their Fig.~16).  In their models the
stellar radiation field is also latitude-dependent, and so impacts
the ionization balance in addition to the varying wind density
profile with latitude.  Those authors are interested mainly in the
inner wind region, where recombination and resonance lines form,
and they do not consider the formation of forbidden lines at large
radius.

Although we could have allowed for a stellar radiation field that
was latitude-dependent, we have chosen to take the stellar radiation
field as isotropic and focus on the large-scale asymptotic properties
of the wind.  But it should be noted that the latitude-dependence
of the stellar radiation field and the latitude-dependent mass flux
will not be independent, and could act to enhance or suppress the
effects presented here, the question of which trend being dependent
on the physics of the mass-loss processes and associated radiation
transport.

\subsection{Model Results}	\label{models}

For computing a selection of model line profiles, we have chosen to focus
on forbidden lines for the element of neon that are of particular interest
for WR~stars.  The WR~winds are selected because they show a number
of strong forbidden emission lines in their IR spectra as revealed by
ground-based and satellite observations (Barlow \etal\ 1988; Willis \etal\
1997; Morris \etal\ 2000; Dessart \etal\ 2000; Ignace \etal\ 2001; Morris,
Crowther, \& Houck 2004), and these are important for determining metal
abundances that can be used to test models of massive star evolution
(e.g., Meynet \& Maeder 2003).  Neon is selected because lines
from multiple stages of ionization of Ne can be observed in the IR,
such as [Ne{\sc ii}] and [Ne{\sc iii}], and possibly [Ne{\sc v}].

Note that our models neglect clumping, which warrants a comment or two.
Although clumping is well-known to affect diagnostics of WR~winds, such
as recombination lines (Hillier 1991) and radio emission (Abbott \etal\
1981; Nugis, Crowther, \& Willis 1998), and there is a growing body
of evidence that clumping also influences the less dense O star winds
(Crowther \etal\ 2002; Bouret, Lanz, \& Hillier 2005; Fullerton, Massa,
\& Prinja 2006), a good theoretical understanding of wind clumping is
lacking, because the numerical calculations are rather complex (although
there have been advances on this topic by Dessart \& Owocki 2003, 2005).
The details of how clumping initiates, where it initiates, how it evolves
in the wind, and how it might vary in response to a latitudinal density
distribution (not to mention the ionization balance) remain more or
less open questions.  For spherical winds a constant clumping factor
is straightforward to include, such as in Dessart \etal\ (2000) or
Smith \& Houck (2005).  Assuming clumping did not vary with radius
or latitude, it could be included in our expressions as a simple
multiplicative correction factor, but we have
chosen to ignore clumping in our models for now, because of the many
questions currently surrounding the details of clumping.

\subsubsection{Ionization of Neon}

\begin{table}
\caption{Atomic Data for Neon	\label{tab1}}
\begin{tabular}{cc}
\hline Recombination Rates & Photo-Ionization Cross-sections$^d$ \\ 
 ($10^{-12}$ cm$^3$ s$^{-1}$) & ($10^{-18}$ cm$^2$) \\ \hline
 ---  & $\sigma_0 = 58.9$ \\
$\alpha_1^a    = 0.283 \times \sqrt{10^4/T_{\rm e}}$ & $\sigma_1 = 14.6$\\
$\alpha_2^a    = 1.71 \; \times \sqrt{10^4/T_{\rm e}}$ & $\sigma_2 = \; 2.7$\\
$\alpha_3^a    = 4.44 \; \times \sqrt{10^4/T_{\rm e}}$ & $\sigma_3 = \; 3.7$\\
$\alpha_4^a    = 9.81 \; \times \sqrt{10^4/T_{\rm e}}$ & $\sigma_4 = 0.54$\\
$\alpha_5^b    = 25.6 \; \times \sqrt{10^4/T_{\rm e}}$ & $\sigma_5 = 0.26$\\
$\alpha_6^b    = 42.7 \; \times \sqrt{10^4/T_{\rm e}}$ & $\sigma_6 = 0.49$\\
$\alpha_7^b    = 65.8 \; \times \sqrt{10^4/T_{\rm e}}$ & $\sigma_7 = 0.22$\\
$\alpha_8^c    = 320 \; \times \sqrt{633/T_{\rm e}}$ & $\sigma_8  = 0.19$\\
$\alpha_9^c    = 616 \; \times \sqrt{327/T_{\rm e}}$ & $\sigma_9  = 0.069$\\
$\alpha_{10}^c = 1085 \times \sqrt{183/T_{\rm e}}$ & --- \\ \hline
\end{tabular}

$^a$ From Osterbrock (1989) \
$^b$ Based on interpolation (see text) 
$^c$ From Verner \& Ferland (1996)
$^d$ From Verner \etal\ (1996) \

\end{table}

We have computed the ionization balance for Ne using the semi-analytic
results of equation~(\ref{eq:Q}).  Table~\ref{tab1} lists the atomic data
used for these computations.  The recombination coefficients $\alpha_{\rm
i}$ where the subscript $i$ refers to recombination from Ne$^{+i}$ to
the next lower ion stage, were taken from Osterbrock (1989) for $i=1-4$.
Values of $\alpha_{\rm i}$ for $i=8-10$ come from Verner \& Ferland
(1996).  The remaining coefficients were computed from an interpolation
of these data.  The photo-ionization cross-sections are values at the
respective ionization edge frequencies and were taken from Verner \etal\
(1996).  Here the subscript refers to photo-ionization from Ne$^{+i}$
to the next higher ion stage.

We have simplified the calculations with two assumptions:  (a) the
stellar radiation field is Planckian and (b) the photo-absorption
cross-sections scales as $\sigma_\nu = \sigma_{\rm i}\,(\nu_{\rm
i}/\nu)^{3}$, where $\sigma_{\rm i}$ and $\nu_{\rm i}$ are values
at the photo-absorption edge.  Although these simplifying assumptions
will change the detailed quantitative results of our models, they
should not impact the qualitative results.  For example, the scaling
for the ion fractions presented in equation~(\ref{eq:scale}) should
still hold.

\begin{figure}
\centerline{\epsfig{file=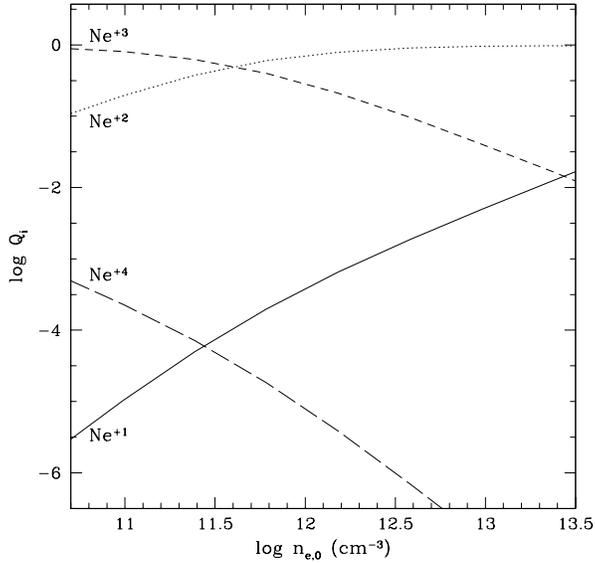,angle=0,width=8cm}}
\caption[]{
A plot of the ion fractions $Q_{\rm i}$ for Ne$^{+i}$ against the base
wind density scale parameter $n_{\rm e,0}$.  Ne {\sc iv} is dominant for
the lower densities in this plot, but Ne {\sc iii} becomes dominant for
$n_{\rm e,0}$ above about $3\times 10^{11}$ cm$^{-3}$.  As described in
the text, these results assume a photoionizing spectrum from a 40,000~K
blackbody.
\label{fig4}}
\end{figure}

An example of the ionization balance is shown in Fig.~\ref{fig4} for
neon plottted against the wind density scaling parameter $n_{\rm e,0}$.
The radiation field is taken to be Planckian with a temperature of
40,000~K, a reasonable effective temperature for WR stars derived from
detailed spectral analyses (e.g., Hamann \& Koesterke 1998).  It is
stressed that the ion fractions are plotted against $n_{\rm e,0}$
for different stars, and do not represent varying ion fractions as a
function of wind density in a single star.  For a given $n_{\rm e,0}$,
these are the ion fractions at every radius.  The range of values for
$n_{\rm e,0}$ are typical for WR winds, and the dominant ion varies
from Ne {\sc iv} at the lower densities to Ne {\sc iii} at the higher
densities.  At this temperature, Ne {\sc ii} is quite trace.
Smith \& Houck (2001) have found that [Ne {\sc ii}] 12.55 microns
appears in the spectra of some late WN and WC stars.  For reference
our models show that for a lower blackbody temperature of 30,000 K,
Ne {\sc ii} becomes the dominant ion above $2\times 10^{13}$ cm$^3$.

The results of Fig.~\ref{fig4} are admittedly a cheat.  For the
dense WR~winds, changing the wind density scale impacts the emergent
radiation field, which has feedback for the ionization balance
in the wind.  So it is not really correct to fix $T_{\rm eff}$ at
40,000~K and allow $n_{\rm e,0}$ to vary independently.  We have
computed ionization fractions based on WR spectral models provided
online\footnote{www.astro.physik.uni-potsdam.de/PoWR.html} by the Potsdam
group for WN and WC stars (Gr\"{a}fener, Koesterke, \& Hamann 2002; Hamann
\& Gr\"{a}fener 2003, 2004).  These treat the radiation field and the wind
density profile consistently.  For a broad sampling of the models provided
in terms of the transformed radius (Schmutz, Hamann, Wessolowski 1989)
and stellar temperature, and using equation~(\ref{eq:Q}), the dominant
ion stage is typically either Ne{\sc ii} or Ne{\sc iii}, both being
relatively strong forbidden lines commonly observed in the IR spectra
of WR stars.  Infrequently, a higher ion stage is seen to be dominant,
such as Ne{\sc v}.

Although it would be more accurate for any given WR~star to use
the stellar radiation data from models like those of the Potsdam
group (even better, the actual computed ionization data, but that
information is not provided at their website), and to have
better atomic data to relax assumption (b) above,
ours is really an exploratory paper to investigate trends that
axisymmetric winds will have for forbidden emission profiles.
It is sufficient at this point to treat the latitude dependence
of the density and ionization balance self-consistently within
the stated assumptions to arrive at valid qualitative conclusions.

\begin{table}
\caption{Model Parameters	\label{tab2}}
\begin{tabular}{clcccc}
\hline Figure-Panel$^a$ & Line Type & Ion & $i$ & $G$ & $m$ \\ \hline
5(a), 6(a) & solid & Ne {\sc ii} & $0^\circ$ & 3 & 5 \\
5(a), 6(a) & dotted & Ne {\sc iii} & $0^\circ$ & 3 & 5 \\
5(a), 6(a) & short dash & Ne {\sc iv} & $0^\circ$ & 3 & 5 \\
5(a), 6(a) & long dash & Ne {\sc v} & $0^\circ$ & 3 & 5 \\
     &           &             &           &   &   \\
5(b), 6(b) & solid & Ne {\sc ii} & $90^\circ$ & 3 & 5 \\
5(b), 6(b) & dotted & Ne {\sc iii} & $90^\circ$ & 3 & 5 \\
5(b), 6(b) & short dash & Ne {\sc iv} & $90^\circ$ & 3 & 5 \\
5(b), 6(b) & long dash & Ne {\sc v} & $90^\circ$ & 3 & 5 \\
     &           &             &           &   &   \\
5(c), 6(c) & solid & Ne {\sc iii} & $30^\circ$ & 1 & 3 \\
5(c), 6(c) & dotted & Ne {\sc iii} & $30^\circ$ & 2 & 3 \\
5(c), 6(c) & short dash & Ne {\sc iii} & $30^\circ$ & 3 & 3 \\
5(c), 6(c) & long dash & Ne {\sc iii} & $30^\circ$ & 9 & 3 \\
     &           &             &           &   &   \\
5(d), 6(d) & solid & Ne {\sc iii} & $60^\circ$ & 1 & 3 \\
5(d), 6(d) & dotted & Ne {\sc iii} & $60^\circ$ & 2 & 3 \\
5(d), 6(d) & short dash & Ne {\sc iii} & $60^\circ$ & 3 & 3 \\
5(d), 6(d) & long dash & Ne {\sc iii} & $60^\circ$ & 9 & 3 \\
     &           &             &           &   &   \\
5(e), 6(e) & solid & Ne {\sc iii} & $60^\circ$ & 3 & 1 \\
5(e), 6(e) & dotted & Ne {\sc iii} & $60^\circ$ & 3 & 3 \\
5(e), 6(e) & short dash & Ne {\sc iii} & $60^\circ$ & 3 & 4 \\
5(e), 6(e) & long dash & Ne {\sc iii} & $60^\circ$ & 3 & 8 \\
     &           &             &           &   &   \\
5(f), 6(f) & solid & Ne {\sc iii} & $0^\circ$ & 3 & 3 \\
5(f), 6(f) & dotted & Ne {\sc iii} & $30^\circ$ & 3 & 3 \\
5(f), 6(f) & short dash & Ne {\sc iii} & $60^\circ$ & 3 & 3 \\
5(f), 6(f) & long dash & Ne {\sc iii} & $90^\circ$ & 3 & 3 \\ \hline
\end{tabular}

\centering{$^a$ Fig.~5 is for the equator denser than the pole;
Fig.~6 is for the pole denser than the equator.  See eqns.~(24) and
(25).}

\end{table}

For an axisymmetric wind, one has $n_{\rm e,0}$ a function of
latitude.  Fig.~\ref{fig4} indicates that changes in $n_{\rm e,0}$
with latitude will lead to changes in the values of the ion fractions,
possibly even in a shift of the dominant ion stage.  So both the
density and the ion fractions will change.  Generally, as the base
density increases, the ion fraction for a stage $i>i_0$ will tend
to drop, thereby somewhat compensating for the effect of increased
density on the emissivity.  On the other hand if $i<i_0$, then
increasing the base density will lead to a higher ion fraction,
thus compounding the effect of an increased density for the emissivity.

\subsubsection{Line Profile Generation}

For an optically thin emission line formed over a large volume, the
total line luminosity from an axisymmetric wind will be independent of
the viewing inclination, and given by the following integration:

\begin{equation}
L_l = 8\pi^2\,\int_{R_*}^\infty\,r^2\,dr\,\int_{-1}^{+1}\,
	j(r,\theta_*)\,d\mu_*,
\end{equation}

\noindent However, the emission profile shapes will depend on viewing
inclination.

We had previously defined stellar spherical coordinates.  We now
introduce stellar Cartesian coordinates $(x_*, y_*, z_*)$ and observer
coordinates $(x, y, z)$.  The observer is located along the positive
$z$-axis.  The viewing inclination angle $i$ is defined as the angle
between $z$ and the stellar symmetry axis $z_*$, thus $i=0^\circ$
is a pole-on view, and $i=90^\circ$ is edge-on.  We also choose
$y=y_*$ without loss of generality.  The observer has spherical
angular coordinates $(\theta, \phi)$.

The emission profile shape is given by a volume-integrated emissivity
for different isovelocity zones.  The isovelocity zones are defined
by 

\begin{equation}
v_{\rm z} = - \vinf \, \cos \theta = -\vinf\,\mu,
\end{equation}

\noindent and so are conical-shaped regions.  Noting that $dv_{\rm z}
= - \vinf \,d\mu$, combined with $d\nu = \nu_0\,dv_{\rm z}/c$ to give
$d\nu= (\vinf/\lambda_0)\,d\mu$, an integral expression for
the profile shape $dL_l/d\nu$ becomes

\begin{equation}
\frac{dL_l}{d\nu} (v_{\rm z}) = 4\pi\,\frac{\lambda_0}{v_\infty}\,
	\int_{R_*}^\infty\,r^2\,dr\,
	\int_{0}^{2\pi}\,j(r,\theta_*)\,d\phi.
\end{equation}

\noindent Evaluation of this integral must generally be carried out numerically
(although see App.~\ref{poleon} for the special pole-on case),
using the spherical trigonometric relation that for $\theta$ and $\phi$ given
for a point in the wind, $\theta_*$ can be found from

\begin{equation}
\cos \theta_* = \cos \theta \,\cos i + \sin\theta\,\sin i\,\cos\phi.
\end{equation}

\noindent Clearly for an arbitrary viewing inclination $i$, a cone of fixed
$\theta$ will cut across stellar latitudes from $\theta_* = \theta-i$
to $\theta+i$, and so a particular isovelocity zone at $\theta$ will
sample different densities and ion fractions.  

\begin{figure*}
\centerline{\epsfig{file=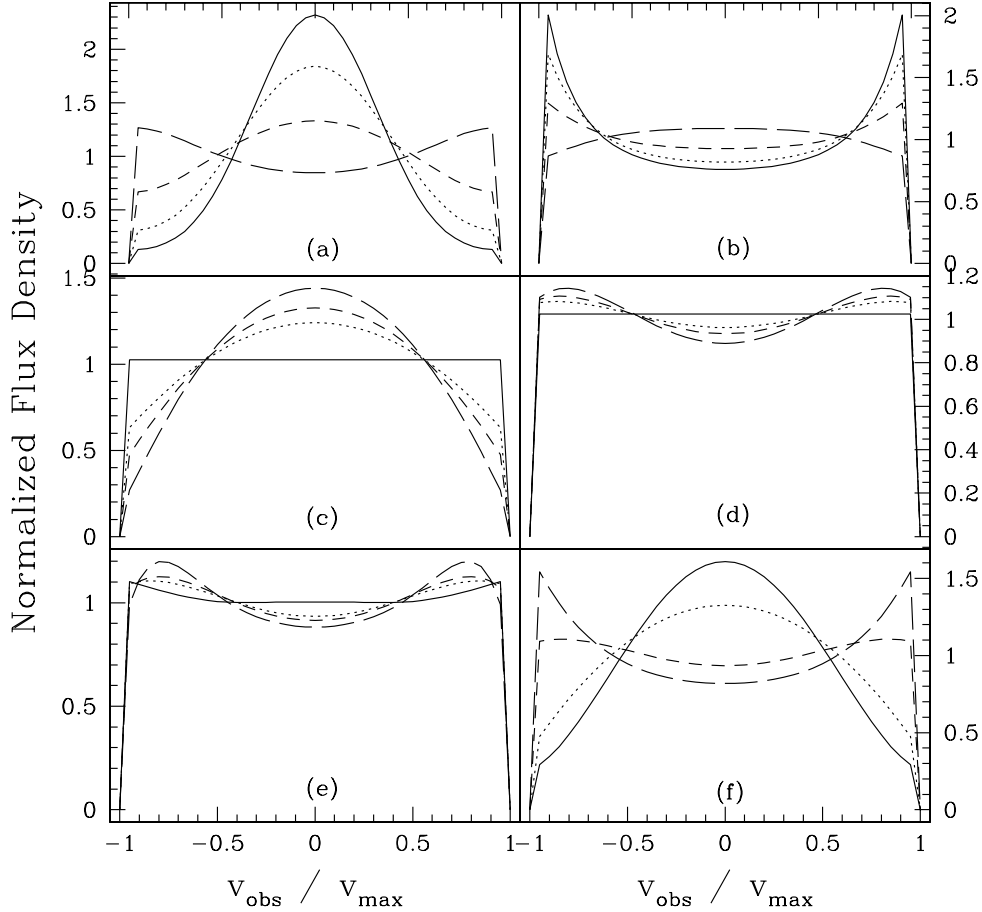,angle=0,width=13cm}}
\caption[]{
Model forbidden line profiles for axisymmetric winds that are more dense
at the equator than the pole.  Table~\ref{tab2} provides a listing of
the model parameters and ions of neon being modeled for each of the line
types in the different panels.  The ordinate is normalized flux density
(such that all of the emission lines have the same equivalent widths),
whereas the abscissa is normalized doppler shift.  Note that the vertical
scale is different for each panel as indicated.  The axisymmetric wind is
taken to have $v_{\rm max}$ the same for all latitudes, and $v_{\rm obs}$
varies from $-v_{\rm max}$ from the left side of a profile to $+v_{\rm
max}$ at the right side of a profile.  Each profile drops to zero line
emission at the extreme velocity shift values.  \label{fig5}}
\end{figure*}

\begin{figure*}
\centerline{\epsfig{file=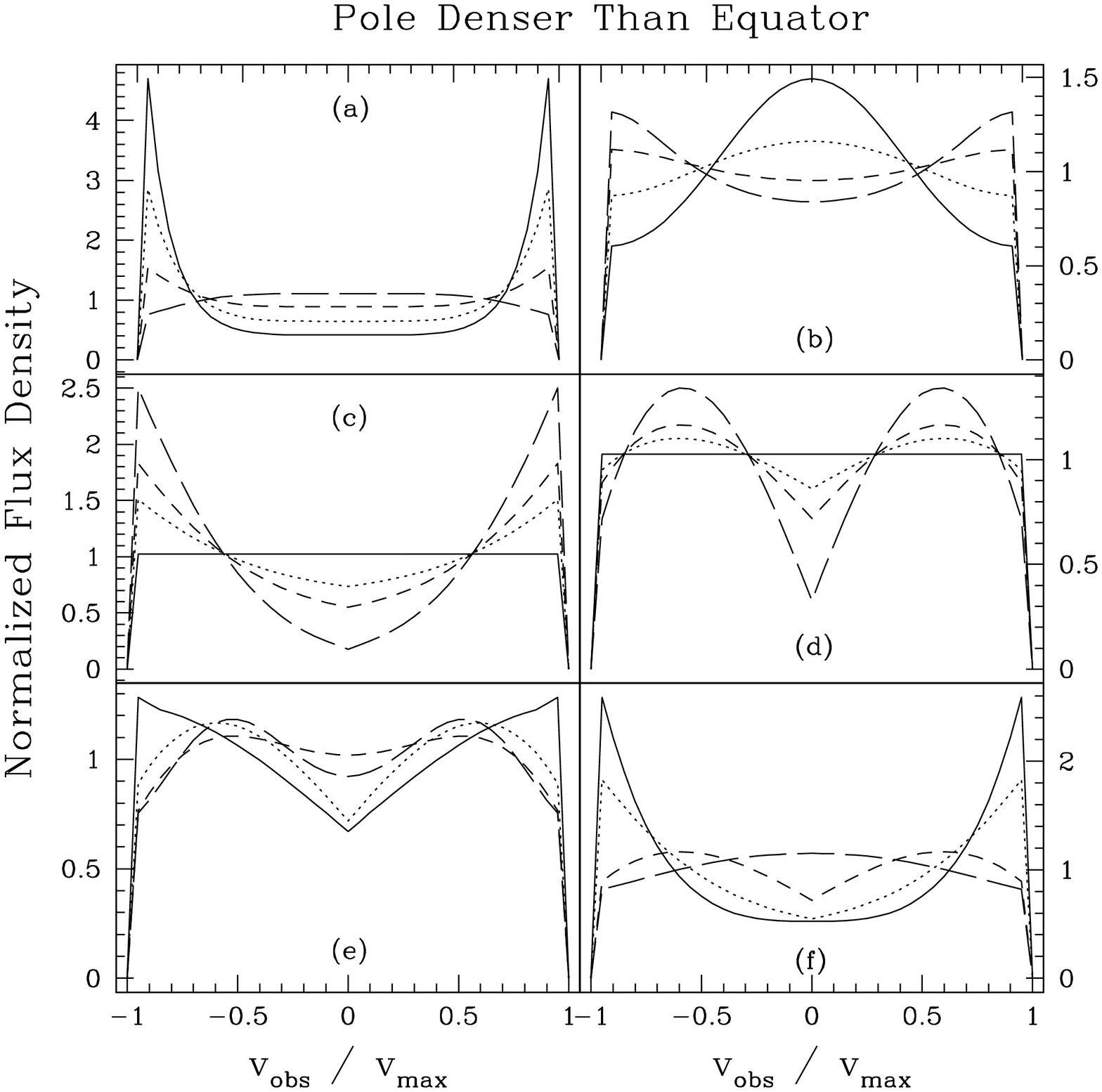,angle=0,width=13cm}}
\caption[]{
As in Fig.~\ref{fig6}, but now for winds that are more dense at the poles
than in the equator.  Again, a key for the different line profiles is to
be found in Table~\ref{tab2}.
\label{fig6}}
\end{figure*}

\subsubsection{Parameter Study}

A number of model forbidden line profile shapes have been computed for
ions of neon, and these are shown in Fig.~\ref{fig5} for dense equatorial
winds and Fig.~\ref{fig6} for dense polar winds.  A key to the different
models displayed there is provided in Table~\ref{tab2}.  The models assume
a wind dominated by He{\sc ii} at large radius with $\mu=2$, $\mu_{\rm
e}=4$, a total wind density scale of $n_0=6\times 10^{12}$ cm$^{-3}$, and
an effective blackbody temperature of $T=40,000$~K.  Figure~\ref{fig4}
shows that Ne {\sc iii} will be dominant, with Ne {\sc iv} down by a
factor of about 10, and other ion stages inferior in abundance by much
larger factors.

These parameters are roughly chosen to match the wind properties of
WR~134 (Hamann \& Koesterke 1998).  The motivation is that WR~134 is
known to have an asymmetric wind (Underhill \etal\ 1990; Vreux \etal\
1992;  Schulte-Ladbeck \etal\ 1992; Harries, Hillier, \& Howarth 1998).
The emission line of [Ca{\sc iv}] 3.21 $\mu$m obtained from {\it ISO}
and discussed in Ignace \etal\ (2001) hints at a double-peaked emission
profile, although the signal-to-noise is not very good in this weak line.
Unpublished spectra of the strong He{\sc ii} 3.09 $\mu$m recombination
emission line in WR~134 clearly shows a well-defined, albeit asymmetric,
double-peaked shape, similar to emission lines shapes
that have been reported by others.

Although the basal wind density is consistent with the wind and star
parameters reported by the Potsdam group, the stellar temperature of
40,000~K was chosen to approximate the correct ionization balance; in fact
$T_*$ reported by Hamann \& Koesterke (1998) is about twice as high, but
the He{\sc ii} ionization edge causes the spectrum to deviate strongly
from Planckian.  As discussed previously, these are simply details in
light of the primary goal of highlighting qualitative profile effects.

There are six panels in Figs. \ref{fig5} and \ref{fig6}, with each
showing several line profiles.  All of the emission profiles are plotted
as normalized flux density against normalized Doppler shift.  The flux
density is related to the specific line luminosity via $dL_l/d\nu = 4\pi
D^2F_\nu$.  For the normalization, the integrated flux density of the
line is used so that each continuum-subtracted line-profile would have
a flat-topped profile of unit height for a spherically symmetric wind.
Note that each panel has a different vertical scale, indicated at the
right or left side.

The panels illustrate the range of profile shapes that can result as
different parameters are varied.  Each panel is labeled (a)--(f) in
the respective figures, and a key for the model parameters with each
profile is provided in Table~\ref{tab2}.  In that table the value of $G$
for Fig.~\ref{fig5} is the ratio of the equatorial to polar densities
at the same radius, but for Fig.~\ref{fig6}, $G$ is the polar to
equatorial density contrast.  In both figures, the top two panels are
for the extreme viewing inclinations of pole-on (left) and equator-on
(right), with a mild density enhancement of $G=3$ and a power exponent
of $m=5$.  For the case of the dense polar wind seen at $i=0^\circ$, the
profiles evolve from strongly double-peaked for Ne {\sc ii} (solid)
to modestly centrally bubbled for Ne {\sc v} (long dash).  Equator-on,
the trend is opposite.  The parameters $G$ and $m$ are fixed, so these
are examples of reversals in the qualitative shape of the profiles
as feedback from the ionization dependence on the density scale with
consequence for the line emissivity.  For this pole-on case, an analytic
solution to the profile shapes and these scaling effects can be derived,
which is provided in Appendix~\ref{poleon}.

The two middle panels are for the fixed dominant ion state of Ne{\sc iii},
with $m=3$ fixed, and $G$ allowed to vary from 1 to 9.  The case $G=1$ is
spherical and so produces the flat-topped profile.  The left middle panel
is for a viewing inclination of $i=30^\circ$, whereas the right middle
panel is for $i=60^\circ$.  At lower left the exponent $m$ is allowed to
vary from 1 through 8 as indicated, with $G=3$ and $i=60^\circ$ fixed
for the Ne{\sc iii} ion.  The panel at lower right overplots profiles
of $m=3$ and $G=3$ for Ne{\sc iii} as the viewing inclination varies.

There are several interesting comments and conclusions to be made from
the parameter study:

\begin{enumerate}

\item The profile distortions from a flat-top shape for the dense polar
wind models as compared to dense equatorial winds appear more severe for
a given value of $G$ and $m$; however, one must interpret this with caution.
Although at fixed $m$, both models have the same opening angle $\Delta \theta$,
this does not translate to the same solid angle.  For example, if
$m=1$ for a polar wind model, then $\Delta \theta = 60^\circ$ (see
eq.~[\ref{eq:angle}]).  The relative solid angle of the dense polar
wind component is $\Delta \Omega / 4\pi = 0.5$.  For a model with a
dense equatorial wind, a structure of the same solid angle using the
density parametrization of equation~(\ref{eq:neq}) would have $\Delta
\theta=30^\circ$ requiring $m=4.8$.  So for example to make a fair
comparison of profiles between Figs.\ \ref{fig5} and \ref{fig6} as $m$
is varied, it is the short dashed line of \ref{fig5}(e) that should be
contrasted with the solid line of \ref{fig6}(e).  In so doing, one finds
that indeed the profile shapes are different between models of dense
equatorial and dense polar winds, but the distortions from flat-top are
comparable in amplitude.

\item Panels (a) and (b) for both Figures shows that quite strong centrally
peaked or double-horned profiles do result for the dominant ion (dotted
line type in these panels) for favorable viewing perspectives of pole-on
or equator-on.  On the other hand, mid-latitude views as shown in panels
(d) and (e) for $i=60^\circ$, statistically a probable viewing angle,
reveal that distortions from flat-top shape are pretty mild, usually
less than 20\% deviations for the range of values in $G$ and $m$
shown.  This is a result of the isovelocity zones, cones in this case,
cutting across a fairly broad range of stellar latitudes and so sampling
both high and low density regions.  The end result is compensation such
that the profile tends toward flat-top, unless the values of $G$ or $m$
become somewhat large.

\item The profiles have been plotted in normalized form to emphasis the
impact of model parameters on line shape.  However, both $G$ and $m$
also impact the total line flux.  If the emissivity were just linear
in density, then the total line emission would be conserved even though
the profile shape were to deviate from flat-top.  The emissivity is only
linear in density for the 'flat' part; it is quadratic in the 'linear'
segment of the contribution function (see Fig.~\ref{fig1}).  As a result,
line flux is not conserved.  Indeed, it may be greater or smaller
(depending on the ion stage) than an equivalent spherical wind of the
same total mass-loss.  For the dominant ion, line flux will be increased
because of deviations from spherical flow.  In general this biases
ratios of line fluxes for deriving relative abundances.  Specifically,
if a line ratio is formed from two ions of different species but with the
same value of $\Delta i$ (and $\Delta i$ is constant with latitude for
both ions), the effect cancels, and ion fraction ratios will be the same
as if the wind were spherically symmetric.  So valid abundance ratios
can be derived even if the wind is axisymmetric, a consequence of the
line being optically thin and stellar occultation effects negligible.
However, this is only true if $\Delta i$ is the same for both lines,
otherwise the ion fraction ratio will deviate from what would have been
obtained from an equivalent spherical wind.

\item The model results clearly show to what extent line profiles will
deviate from flat-top as model parameters are varied, but what can
be concluded about the wind structure given an observed line profile?
Without some {\it a priori} knowledge, there are qualitative degeneracies
which combined with limited signal-to-noise data can make the inverse
problem challenging.  For example, changing ion stage ($\Delta i$) has
effects that are similar to changing $G$ and/or $m$, even toward
making it ambiguous as to whether the wind is denser at the poles or
the equator.  Normally, based on the spectral properties, one does
have an idea about the dominant ion stage.  

If the ion state is roughly known, then increasing $G$ and $m$ tends
to make the profiles display stronger deviations from a flat-top
shape.  Both the total line flux and the amplitude of the features
(central peak or double-horns) must be matched simultaneously as $G$
and $m$ vary.  The width of these features must also be matched, which
may discriminate between dense polar versus dense equatorial models.
The viewing inclination also influences the sharpness of the features
as well as their location within the profile.  Additional data, such as
nebular morphology, polarimetry, or possibly even rotational $v \sin i$
values, may help to limit the range of viewing inclination.  In practice,
if the viewing inclination is intermediate of pole-on and edge-on (as
it likely will be), and the wind distortion is relatively mild, it may
be hard or impossible to confidently determine the wind distortion 
parameters unless both the signal-to-noise and spectral resolution are
quite high.

\end{enumerate}

\section{Discussion}

The objective of this study has been to consider how forbidden emission
line profiles might be used as a diagnostic of the large-scale structure
of stellar winds that deviate from spherically symmetric outflow.
To this end, we have introduced parametric expressions for dense polar
or dense equatorial axisymmetric winds.  The stellar radiation field and
the wind terminal speed have both been treated as isotropic, but these
are assumptions that should be relaxed in producing synthetic line profiles
for a particular model wind.  If the mass-loss from the star varies with
latitude, it is to be expected that the terminal speed will also vary,
possibly correlating or anti-correlating with the mass-loss rate depending
on the specifics of the wind-driving physics.  In combination with the
effects of the ionization balance, deviations of the line profile shape
from flat-top may be suppressed or enhanced compared to the results
presented here.  Relevant are the effects of clumping and velocity
dispersions, such as those investigated by Runacres \& Owocki (2002)
for an O star model at large radius in the flow (100's of stellar radii).
Exactly how such properties will vary will global asymmetries is unclear.
However, for a spherical wind, if such effects persist over the scale
$r_{\rm c}$, they should be detectable in the form of sloping, in
contrast to more nearly vertical, high velocity wings to the forbidden
line profile shape.  For aspherical winds, the expressions in this paper
for thin forbidden lines can be modified to allow for a range of flow
speeds and densities in a ``cell-by-cell'' approach to doing the numerical
volume integration for the emission profile.

A particular focus of this work has been to highlight the coupling
between density and ion stage which are coupled in latitude.  These can
act through the line emissivity to exacerbate or depress deviations
of the line profile shape from a flat-top.  Including the effects of
a latitude-dependent radiation field and terminal speed may modify the
conclusions that we have drawn from our assumed model, but the overall
principle that resolved forbidden emission lines can be excellent
tracers of deviations from spherical flow remains robust.
At the same time, Figs.~\ref{fig5} and \ref{fig6} do indicate that for
winds with modest distortions from spherical as viewed somewhat more
equator-on (the case $i=60^\circ$) than pole-on ($i=30^\circ$), the line
profiles of the dominant stage show modest deviations from a flat-top.
Combined with instrumental smearing and finite signal-to-noise, such
profiles might pass as essentially flat-top.  

As an observational strategy, it would be best to obtain forbidden
emission line data for different ion stages within the same atom.  This is
not too common however, as ion stages subordinate to the dominant one
can have much smaller ion fraction values and produce only weak lines.
The lines of [Ne{\sc ii}] 12.86 $\mu$m and [Ne{\sc iii}] 15.56~$\mu$m are
an example of a pair of forbidden lines in adjacent ion stages that can
both be seen in WR star winds (see App.~\ref{2lvl} for a discussion
of the applicability of the two-level atom approximation for Ne{\sc
iii}).  {\it ISO} observed WR stars with high spectral resolution at
the appropriate wavelengths, but only a few such stars were observed.
One example showing both forbidden lines is WR~11 (van der Hucht \etal\
1996), but this is a colliding wind binary system.  Both lines were
also seen in the luminous blue variable, P Cygni (Lamers \etal\ 1996).
Already {\it Spitzer} has observed a number of WR stars with the IRS;
unfortunately, the IRS generally has inadequate spectral resolution to
analyze the line profile shapes to apply the results of this paper.
An exception to this might be the extremely fast, but also extremely
rare, winds of the oxygen-rich WO subtype ((Barlow \& Hummer 1982;
Kingsburgh \etal\ 1995).  These winds sport terminal speeds upwards of
3000 km/s (e.g., Drew \etal\ 2004).  Forbidden lines observed at high
resolution with the IRS for the WO stars may be suitable to determine
or place limits (subject to the signal-to-noise) on the geometry of the
asymptotic structure of these winds.  These are hotter stars, so instead
of [Ne {\sc iii}], perhaps IR lines of [Ne {\sc v}] will be present.

\section*{Acknowledgements}

This research was conducted in part through a summer REU program,
for which Adam Brimeyer expresses grateful appreciation to the SARA REU
program and the National Science Foundation.  This research was supported
in part by a grant award to the Florida Institute of Technology by the
NSF (AST-0097616).  The authors also thank Drs Gary Henson, and Joe
Cassinelli, and Martin Hendry for helpful comments pertaining to this
study.  Ignace is particularly grateful to Ken Gayley for discussions in
the early part of this study.  The authors are also very appreciative of
an anonymous referee who made several helpful comments that significantly
clarified the presentation of results.

\appendix

\section{Analytic Line Profiles for a Pole-On Inclination}	\label{poleon}

\begin{figure*}
\centerline{\epsfig{file=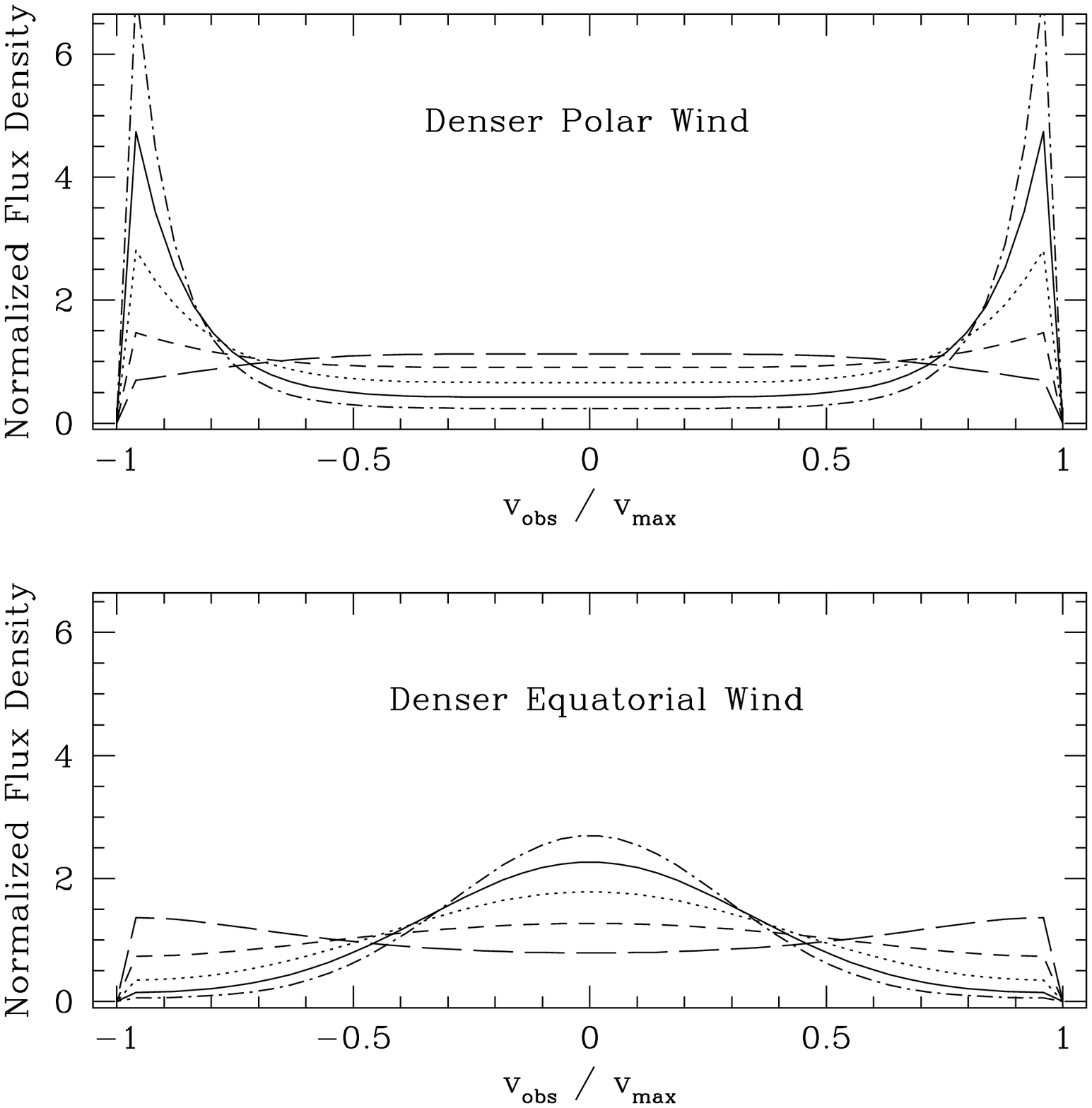,angle=0,width=13cm}}
\caption[]{
Analytic solutions for forbidden line profile shapes as seen pole-on.
Upper is for a dense polar wind, and lower is for a dense equatorial wind.
For the dense polar wind of the upper panel, profiles progressively
shift from a strongly double-horned emission shape with $\Delta i=-2$
to one that is mildly centrally peaked at $\Delta i = +2$.  The exact
opposite progression occurs for a dense equatorial wind in the lower
panel.  Values of $m=5$ and $G=3$ were adopted, and the profiles are
normalized so that a spherical wind would produce a profile of unit
height in this figure.
\label{fig7}}
\end{figure*}

The latitude dependence of the axisymmetric density distribution is
described by

\begin{equation}
f(\theta_*) \equiv n(r,\theta_*)/n_{\rm sph}(r) = H^{-1}\,\left[ 1 +
	(G-1)\,g(\theta_*)\right],
\end{equation}

\noindent where $H$ is a normalization set by eq.~(\ref{eq:norm}),
and $g(\theta_*) = \cos^m \theta_*$ for a dense polar wind or
$\sin^m \theta_*$ for a dense equatorial wind.

In the case of a pole-on viewing inclination with $i=0^\circ$, one has
that $\theta=\theta_*$.  The emission line profile shape is then determined
by

\begin{eqnarray}
\frac{dL_l}{d\nu} & \propto & \int_{R_*}^\infty\, n_2(r)\,r^2 \,dr \\
 & \propto & \int_{R_*}^\infty\, \frac{n_{\rm e}\,Q_{\rm i}}{1+n_{\rm c}/n_{\rm e}}\, r^2 dr 
\end{eqnarray}

\noindent Using eq.~(\ref{eq:scale}) and defining $\Delta i = i-i_0$, we
have that

\begin{eqnarray}
\frac{dL_l}{d\nu} & \propto & \int_{R_*}^\infty\,\frac{f^{1-\Delta i}}{1+f\,r^2/r_{\rm c}^2} \, dr\\
 & = & \int_0^1\,f^{1-\Delta i}\,\left(u^2+f^{-1}\,u_{\rm c}^2\right)^{-1}\,du 
\end{eqnarray}

\noindent where the last line involves a substitution of $u=R_*/r$
and the lower limit to the integral assumes $u_{\rm c} = R_*/r_{\rm c} \ll 1$.
This last integral has an analytic solution:

\begin{equation}
\frac{L_l}{d\nu} \propto [f(\theta_*)]^{1.5-\Delta i}.
	\label{eq:pole}
\end{equation}

\noindent Since $\theta=\theta_*$, and the observed doppler shifted
velocity in the profile is $v_{\rm z} = v_{\rm max} \, \cos \theta$,
one has that $\cos \theta_* = v_{\rm z} / v_{\rm max}$, that can be
substituted in for $g(\theta_*)$ to derive an analytic function for the
emission profile shape.

Example profiles based on equation~(\ref{eq:pole}) are shown in
Fig.~\ref{fig7}.  The upper panel is for a dense polar wind, and the
lower one for a dense equatorial wind.  Similar to Figs.~\ref{fig5} and
\ref{fig6}, the line types are for $\Delta i = -1$ (solid), $0$ (dotted),
$+1$ (short dash), $+2$ (long dash).  In those Figures, [Ne {\sc iii}]
was the dominant ion stage for neon, but here Fig.~\ref{fig7} is for
any metal species.  Unlike Figs.~\ref{fig5} and \ref{fig6}, a profile for
$\Delta i = -2$ has been computed for Fig.~\ref{fig7}, and is shown as
dot-dash line type.  In all these cases, values of $G=3$ and $m=5$ were used.
The profiles are normalized by their total emission.

In the case of the dense polar wind as viewed pole-on, the
profiles for the dominant ion stage and lower stages are substantially
double-peaked, because $f$ is large along the poles.  Even though $G$
has a modest value, these profiles depend on $f$ in powers of 1.5, 2.5,
and 3.5, so that strong double-horned profiles still result.  In contrast,
the profiles at $\Delta i = +1$ and $+2$ show a slight double-horned
structure in the first case and a centrally bubbled one in the second.
Both of these appear nearly flat-topped owing to the weak dependence on
$f$ at $f^{-0.5}$ and $f^{+0.5}$, respectively.

In contrast, the profiles for the dense equatorial flows (lower panel) are
all centrally peaked, except for one that shows a modest double-horned
morphology.  Here the denser flow is mostly in the plane of the sky
corresponding to observed velocity shifts around line center.  For the
dominant ion stage and lower, this leads to modest or strong central
peaks relative to a flat-topped profile and substantially depressed
emission wings.  Interestingly, the depressed wings could make the
determination of the true line width problematic owing to uncertainties
in the continuum placement for modest signal-to-noise data.  Once again,
as seen in the upper panel, the profiles for 1 and 2 stages above the
dominant ion approach a somewhat more flat-top appearance, and switches
from centrally peaked emission for $\Delta i = +1$ to the double-horned
appearance for $\Delta i=+2$.

\section{Ne{\sc iii} and the Two-Level Atom Approximation}	\label{2lvl}

The derivations of this paper are predicated on the
assumption of a two-level atom.  A forbidden line of considerable
interest from WR~stars is the [Ne{\sc iii}] 15.55 $\mu$m line
($^3$P$_1$ - $^3$P$_0$), which is seen to be strong in a number of WR
stars from observations by {\it ISO} (Morris \etal\ 2000; Dessart \etal\ 2000)
and {\it Spitzer} (Morris \etal\ 2004; Cassinelli, private comm.).  However,
Ne{\sc iii} has five low-lying energy levels, that are (in order of
increasing energy): $^3$P$_0$, $^3$P$_1$, $^3$P$_2$, $^1$D$_2$, and
$^1$S$_0$.  The question is then, how bad is the two-level approximation
for Ne{\sc iii}?

\begin{figure}
\centerline{\epsfig{file=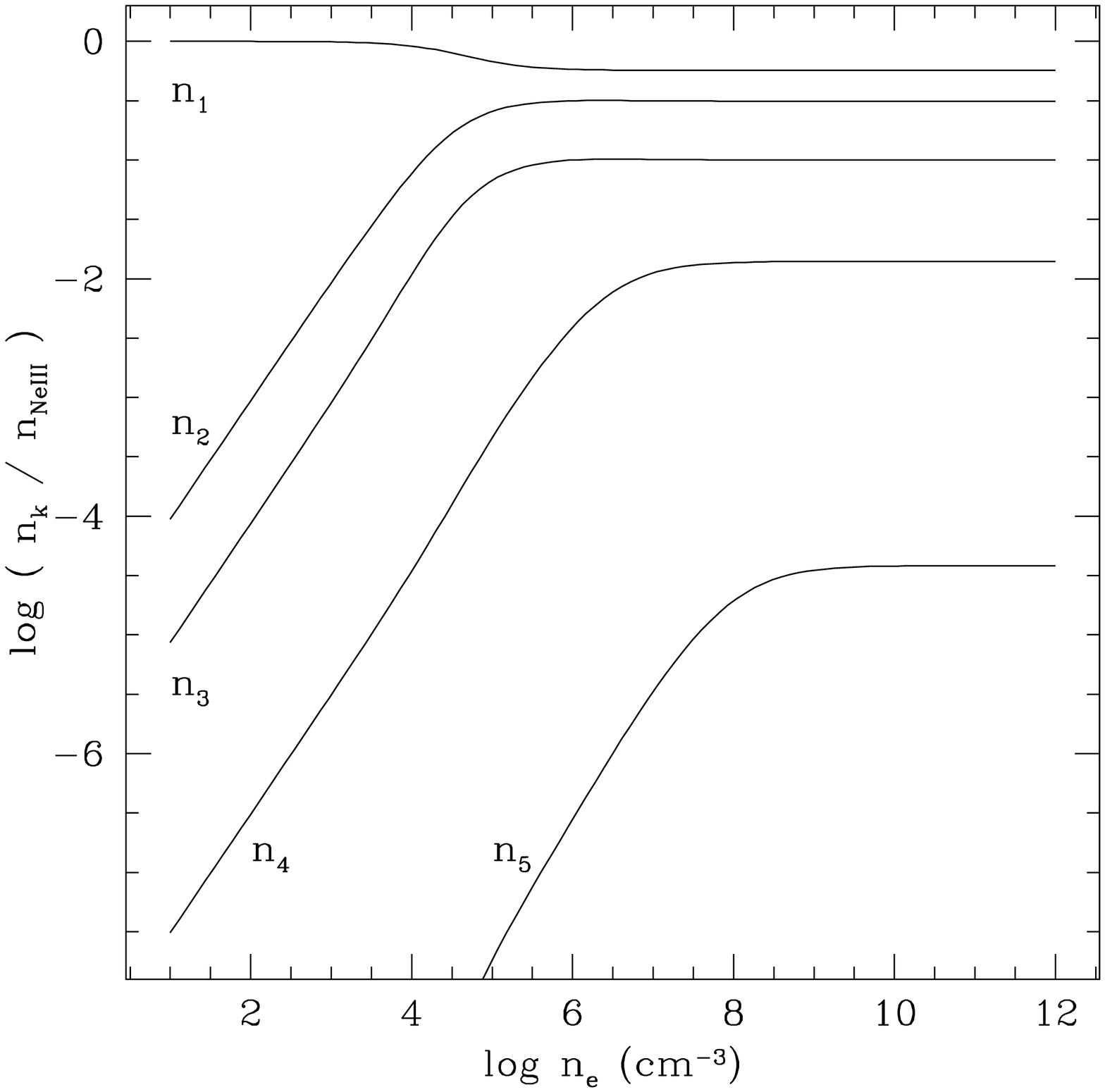,angle=0,width=8cm}}
\caption[]{
A plot of normalized level population number densities 
for the five lowest energy levels
in Ne{\sc iii}.  Assuming highly NLTE conditions, the population in
all levels greater than these five are taken to be negligible.
The transition $2-1$ corresponds to the forbidden line of [Ne{\sc iii}]
15.55 $\mu$m, and this figure shows in large part, this atom can
be reasonably approximated as a two-level atom for conditions relevant
to WR~star winds.
\label{fig8}}
\end{figure}

Population number densities for these five levels, denoted 1-5 in
the order listed above, have been computed, and the results are
displayed in Fig.~\ref{fig8}.  To derive the populations, the
equations of statistical equilibrium were solved allowing for
collisional excitation and de-excitation, and radiative decay.
Being interested primarily in the large radius conditions, the
electron temperature was taken to be $T_{\rm e} = 10,000$ K.  Atomic
data were taken from Pradhan \& Peng (1994).  Absorption and stimulated
emission were included, assuming a blackbody radiation spectrum,
but as expected, these were found to have no impact on the level
populations.

Fig.~\ref{fig8} shows that $n_1 + n_2 \approx n$(Ne {\sc iii})
is reasonably accurate, with an error not exceeding 10\% at densities
higher than the critical density, and for densities near and below
the critical value, the error becomes negligibly small as $n_1 \rightarrow
n$(Ne {\sc iii}).  This conclusion suggests that if Ne {\sc
iii} is the dominant ion, then neon abundances can safely be derived
from the 15.55 $\mu$m line in the two-level atom approximation,
with a correction that is probably comparable or less than measurement
errors associated with continuum placement for determining the line
flux or errors in the distance measurement of the star or of other
fundamental wind parameters (e.g., the mass-loss rate).

\end{document}